\documentstyle[12pt]{article}
\begin{document}
\begin{center}
{\large \bf 
Power Law Distributions for Stock Prices in Financial Markets \\

\vspace*{.5in}

\normalsize 
Kyungsik Kim$^{*}$, S. M. Yoon$^{a}$, K. H. Chang$^{b}$ 

\vspace*{.3in}

{\em 
Department of Physics, Pukyong National University,\\
Pusan 608-737, Korea\\
$^{a}$Division of Economics, Pukyong National University,\\
 Pusan 608-737, Korea \\ 
$^{b}$ Forecast Research Laboratory, Meteorological Research \\ 
Institute, KMA, Seoul 156-720, South Korea } \\
}
\end{center}

\hfill\\
%
%
%
\baselineskip 24pt

We study the rank distribution, the cumulative probability, and the probability density of 
returns of stock prices of listed firms traded in four stock markets.
We find that the rank distribution and the cumulative 
probability of stock prices traded in are consistent approximately
with the Zipf's law or a power law. It is also obtained that the 
probability density of normalized returns for listed stocks almost has the form of 
the exponential function. Our results are compared with those of other numerical calculations.

\vskip 10mm 
\hfill
%
\hfill\\
$^{*}$$E-mail$ $address$: kskim@pknu.ac.kr;
Tel.: +82-51-620-6354; Fax: +82-51-611-6357 \\   

\newpage


The investigation of scaling relations in the Zipf's and Pareto's laws has recently received
considerable attention. Until present, of the many outstanding topics, the main concentration 
is the distribution of the personal income, the distribution of the company size and income, 
the scaling relation of company's size fluctuations and the disribution of the city size.
We believe that these studies are both of much interest and of practical relevance in estimating 
the dynamical behavior of statistical quantities in financial markets. Particularly, 
on the basis of novel statistical methods and approaches of economics, we also expect 
that there has led to a better insight for understanding scaling properties from the result 
of many similar investigations. More than one hundred years ago Pareto $[1]$ investigated the income 
and wealth distributions described by the characteristic feature of a national economics. 
Gini also showed the income distribution approximated by a power law with the non-universal exponent $[2]$,
while Gibrat reported a log-normal distribution of income based on multiplicative random processes $[3]$.
In 1960 Mandelbrot $[4]$ reported that the distribution of incomes scales as a power law, 
and Montroll and Schlesinger $[5]$ showed that the income distribution of high-income group shows a
power law while that of low-income group follows a log-normal distribution $[5]$. Stanley et al. $[6]$ 
recently found that American company size distribution is closer to the log-normal law. 
The Japanese and international company databases are analyzed and reported by Okuyama and Takayasu $[7]$ 
that the distribution function of annual income of companies shows a power law distribution consistent with 
the Zipf's law. Furthermore, many researchers have attempted to compute the distribution functions 
in scientific fields such as the voting process $[8]$, the firm bankruptcy $[9]$, the investment strategy $[10]$, 
the expressed genes $[11,12]$, and the earthquake $[13]$. 

As the first step, it may be of importance to elucidate the scaling properties and statistical methods 
for stock prices. Our purpose in this paper is to find the scaling relation for stock prices of
companies traded on the Korea stock exchange (KSE), Korea securities dealers automated quotations (KOSDAQ), New York stock exchange
(NYSE), and Tokyo stock exchange (TSE). We mainly calculate the cumulative probability and probability density of price returns.   

First of all, we analyze the database of listed companies in four foreign stock markets. The rank for asset $R (p)$ is represented in terms of 
\begin{equation}
\log R (p ) =  \log b + \alpha \log p ,
\label{eq:a1}
\end{equation}
where $a$ and $b$ denote, respectively, the scaling exponent and the parameter, and $R(p)$ denotes the rank of companies in 
the KSE, KOSDAQ, NYSE, and TSE. Eq. $(1)$ is satisfied with the Zipf's law in the case of $ \alpha = -1$. 

To investigate the database both of the KSE listed $668$ Korean companies and of the KOSDAQ Stock
market listed $620$ Korean companies for one-day period ($27$th December of $2002$), Fig. $1$ and Fig. $2$
show, respectively, the distribution of the rank for the KSE and the KOSDAQ. These ranks 
follows the Zipf's law with exponent $\alpha = -1.00$ (KSE) and $-1.31$ (KOSDAQ). 
The estimated rank for NYSE is $\alpha = -3.50$, consistent with the power law as plotted in Fig. $3$,
from the database of the NYSE listed $3,314$ companies of USA for one-day period ($31$st October of $2000$).
Finally, we obtain that the estimated rank for the TSE is $ \alpha = -0.83$ (Fig. $4$), consistent approximately with the Zipf's law, 
from the database of the TSE listed $2,010$ Japanese companies for one-day period ($1$st October of $2000$). However, 
we find from our result that the rank for the KSE, KOSDAQ and TSE follows the Zipf's law, while that for the NYSE shows the power law.

Next the cumulative probability of a stock price is defined as 
\begin{equation}
  P(\geq p) \propto p^{\beta},
\label{eq:b2}
\end{equation}
Here $ \beta $ is the scaling exponent for the cumulative probabilities of the KSE, 
KOSDAQ, NYSE, and TSE. It is found that cumulative probabilities follow a power law 
approximately with scaling exponent $ \beta  = -1.23$ (KSE) and $-1.45$ (KOSDAQ) from Figs. $5$ and $6$. 
The slopes of the cumulative probability of the NYSE and the TSE are, respectively, 
$ \beta = -2.58$ and $-0.83$, as shown in Figs. $7$ and $8$. Hence the cumulative probability of 
the NYSE only follows a power law, but that of other three assets shows the Zipf's law approximately. 
For income distributions, it is really known that Japanese income distribution $[7]$ is closer to
Zipf's law, while a power law holds in the case of USA $[6]$ and Korea $[14]$.

Let $p(t)$ be a stock price at time $t$. We define the price return $ r_{\tau } (t )$, i.e.
the ratio of the stock price of successive times, as follows:
\begin{equation}
r_{\tau } (t ) =  ln \frac{P(t + \tau )} { P(t )} ,
\label{eq:c3}
\end{equation}
where $ \tau $ is the time lag. We introduce the probability density of price returns
as a function of the normalized price return $ r^{*} =(r- <r>)/ \sigma $, where $<r>$ and $ \sigma $ denote, respectively, 
the mean value and the standard deviation of $ r(t)$. It is also found that the probability density of price returns 
almost has the form of an exponential function as
\begin{equation}
P(r^{*} )  \propto \exp (\kappa r^{*} ),
\label{eq:d4}
\end{equation}
where $ \kappa $ is a proportional coefficient. Fig. $9$ depicts the probability density of price returns for KSE 
as a function of the normalized price return with $ \kappa  = -0.57$. 
The propertional coefficients for the KOSDAQ, the NYSE, and the TSE are, respectively, $ \kappa = -0.72$, $ -0.71$, and $-0.65$,
as shown in Figs. $10$-$12$.  

In conclusion, we have discussed the rank distribution function, the cumulative probability, 
and the probability density of price return for four assets, i.e., the KSE, KOSDAQ, NYSE, 
and TSE in financial markets. From our results, the four assets 
investigated so far are related with a power law or Zipf's law.   
It is in particular found that the rank for the KSE, KOSDAQ and TSE follows the Zipf's law, 
while that for the NYSE shows the power law. The cumulative probability of the NYSE is consistent 
with a power law, significantly different from Zipf's law in the KSE, KOSDAQ, and TSE $[6,7]$.
It is also showed that the probability density of normalized price returns almost scales 
as an exponential function, and these studies would of great value 
in characterizing and categorizing the scaling relations in the Zipf's and Pareto's laws 
in other foreign financial markets. In future, we will extensively investigate the tick
data of other stock prices and compare in detail with calculations performed in other nations.\\

\newpage
\vskip 10mm
\begin{center}
{\bf FIGURE  CAPTIONS}
\end{center}

\vspace {5mm}

\noindent
Fig. $1$.  Distribution of the rank for KSE, where the least-squares fit gives 
the power law with exponent $ \alpha = -1.00$.  The solid line shows the Gaussian distribution
$A \exp (-cp^{2} )$ with $A=560$ and $c=4.0 \times 10^{-8}$.
\vspace {10mm}

\noindent
Fig. $2$.  Distribution of the rank for KOSDAQ which scaled as a 
power law with exponent $ \alpha = -1.31$.  The solid line shows the Gaussian distribution
$A \exp (-cp^{2} )$ with $A=620$ and $c=2.5 \times 10^{-8}$.
\vspace {10mm}

\noindent
Fig. $3$.  Distribution of the rank for NYSE which scaled as a 
power law with exponent $ \alpha = -3.50$.  The solid line shows the Gaussian distribution
$A \exp (-cp^{2} )$ with $A=3.3\times 10^{3}$ and $c=1.5 \times 10^{-3}$.
\vspace {10mm}

\noindent
Fig. $4$.  Distribution of the rank for NYSE which scaled as a 
power law with exponent $ \alpha = -0.83$.  The solid line shows the Gaussian distribution
$A \exp (-cp^{2} )$ with $A=2.0\times 10^{3}$ and $c=7.0 \times 10^{-6}$.
\vspace {10mm}

\noindent
Fig. $5$.  Plot of cumulative probability of stocks traded on the KSE, where the slope of the dot line 
is $ \beta = -1.23$.

\vspace {10mm}

\noindent
Fig. $6$.  Plot of cumulative probability of stocks traded on the KOSDAQ, where the slope of the dot line 
is $ \beta = -1.45$.

\vspace {10mm}

\noindent
Fig. $7$.  Plot of cumulative probability of stocks traded on the NYSE, where the slope of the dot line 
is $ \beta = -2.58$.

\vspace {10mm}

\noindent
Fig. $8$.  Plot of cumulative probability of stocks traded on the TSE, where the slope of the dot line 
is $ \beta = -0.83$.

\vspace {10mm}

\noindent
Fig. $9$.  Probability density of price returns for stocks traded on the KOSPI as a function of 
the normalized price return with $ \kappa = -0.57$. The solid line shows the Gaussian distribution 
$ B \exp (- b r^{*} )$ with $ B=0.3$ and $ \sigma=1.0$.
\vspace {10mm}

\noindent
Fig. $10$.  Probability density of price returns for stocks traded on the KOSDAQ as a function of 
the normalized price return with $ \kappa = -0.72$. The solid line shows the Gaussian distribution 
$ B \exp (- b r^{*} )$ with $ B=0.4$ and $ \sigma=0.58$.
\vspace {10mm}

\noindent
Fig. $11$.  Probability density of price returns for stocks traded on the KSE as a function of 
the normalized price return with $ \kappa = -0.71$. The solid line shows the Gaussian distribution 
$ B \exp (- b r^{*} )$ with $ B=0.44$ and $ \sigma=0.66$.
\vspace {10mm}

\noindent
Fig. $12$.  Probability density of price returns for stocks traded on the KSE as a function of 
the normalized price return with $ \kappa = -0.65$. The solid line shows the Gaussian distribution 
$ B \exp (- b r^{*} )$ with $ B=0.35$ and $ \sigma=0.61$.
\vspace {10mm}


\begin{thebibliography}{}

%
\bibitem{Par1} V. Pareto, Le Cours d$^{'}$$\acute{E}$conomie Politique, Macmilan, London, 1897.
\bibitem{Gin2} C. Gini, Biblioteca delli'economista {\bf 20} (1922) 77.
\bibitem{Gib3} R. Gibrat, Les Inegalites Economiques, Sirey, Paris, 1931.
\bibitem{Man4} B. B. Mandelbrot, Int. Econ. Rev. {\bf 1} (1960) 79. 
\bibitem{Mon5} E. W. Montroll and M. F. Schlesinger, J. Stat. Phys. {\bf 32} (1983) 209. 
\bibitem{Sta6} M. H. R. Stanley, L. A. N. Amaral, S. V. Buldyrev, S. Havlin, H. Leschhron, 
P. Maass, M. A. Salinger and H. E. Stanley, Nature {\bf 397} (1996) 804; L. A. N. Amaral, S. V. Buldyrev, S. Havlin,
H. Leschhron, P. Maass, M. A. Salinger, H. E. Stanley and M. H. R. Stanley, J. Phys. I {\bf 7} (1997) 621.
\bibitem{Oku7} K. Okuyama, M. Takayasu and H. Takayasu, Physica A {\bf 269}(1999) 125; H. Takayasu and K. Okuyama, 
Fractals {\bf 6} (1998) 67; K. Kawamura and N. Hatano, J. Phys. Soc. Jpn {\bf 71}, 1211 (2002).
\bibitem{Yen8} M. L. Lyra, U. M. S. Costa, R. N. Filho and J. S. Andrade Jr., cond-mat/0211560. 
\bibitem{Fu9} Y. Fujiwara, C. D. Guilmi, H. Aoyama, and M. Gallegati and W. Souma, Physica A {\bf 335} (2004) 197.
\bibitem{Fuj10} Y. Y. Fujiwara, cond-mat/0310062.
\bibitem{Aus11}  M. Ausloos and Ph. Bronlet, Physica A {\bf 324} (2003) 30; C. Furusawa and K. Kaneko, physics/0209103; 
S. P. Li, K.- l. Ng and M. C. Chung, Physica A {\bf 321} (2003) 189.
\bibitem{Abe12} S. Abe and N. Suzuki, cond-mat/0208344.
\bibitem{Hei13} K. Kim, S.-M. Yoon, C. C. Lee and K. H. Chang, to be submitted.
\bibitem{Mue14} K. Kim and S.-M. Yoon, cond-mat/0403161.
 
%
\end{thebibliography}
\end{document}